\documentclass[]{AO4ELT}  

\usepackage{microtype}
\usepackage{biblatex}
\usepackage{amsmath,amsfonts,amssymb}
\usepackage{graphicx}
\usepackage{caption}
\usepackage{pst-all} 
\usepackage[colorlinks=true, allcolors=blue]{hyperref}
\makeatletter         
\def\@maketitle{
\includegraphics[width = 170mm]{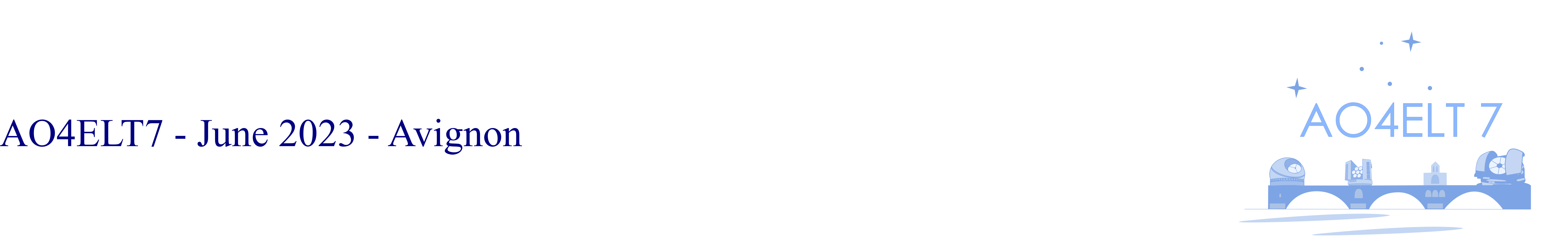}\\[8ex]
\begin{center}
{\Huge \bfseries \sffamily \@title }\\[4ex] 
{\Large  \@author}\\[4ex] 
\@date
\end{center}}

\title{SAXO+ upgrade : second stage AO system end-to-end numerical simulations}

\author[a]{Charles Goulas}
\author[a]{Fabrice Vidal}
\author[a]{Raphaël Galicher}
\author[a]{Johan Mazoyer}
\author[a]{Florian Ferreira}
\author[a]{Arnaud Sevin}
\author[a]{Anthony Boccaletti}
\author[a]{Éric Gendron}
\author[b]{Clémentine Béchet}
\author[b]{Michel Tallon}
\author[b]{Maud Langlois}
\author[c]{Caroline Kulcsár}
\author[c]{Henri-François Raynaud}
\author[c]{Nicolas Galland}
\author[d]{Laura Schreiber}
\author[e]{Gaël Chauvin}
\author[f]{Julien Milli}

\affil[a]{LESIA, Observatoire de Paris-PSL, Sorbonne Université, Université Paris-Cité, CNRS, 5 place Jules Janssen, 92195 Meudon, France}
\affil[b]{CRAL, Université Claude Bernard Lyon 1, ENS de Lyon, CNRS, 9 avenue Charles André, 69561 Saint-Genis-Laval, France}
\affil[c]{IOGS, Université Paris-Saclay, Laboratoire Charles Fabry, CNRS, 2 avenue Augustin Fresnel 91127 Palaiseau, France}
\affil[d]{INAF, Osservatorio di Astrofisica e Scienza dello Spazio di Bologna, Via Gobetti 93/3, 40129, Bologna, Italy}
\affil[e]{Laboratoire J.-L. Lagrange, OCA, Université Côte d'Azur, CNRS, Bd de l’Observatoire, 06304 Nice, France}
\affil[f]{IPAG, OSUG, Univsersité Grenoble Alpes, INSU, CNRS, 414 rue de la Piscine, 38400 St-Martin d’Hères, France}

\authorinfo{E-mail : charles.goulas@obspm.fr}

\begin{document} 
\maketitle
\begin{abstract}
SAXO+ is a proposed upgrade to SAXO, the AO system of the SPHERE instrument on the ESO Very Large Telescope. It will improve the capabilities of the instrument for the detection and characterization of young giant planets. It includes a second stage adaptive optics system composed of a dedicated near-infrared wavefront sensor and a deformable mirror. This second stage will remove the residual wavefront errors left by the current primary AO loop (SAXO). This paper focuses on the numerical simulations of the second stage (SAXO+) and concludes on the impact of the main AO parameters used to build the design strategy. Using an end-to-end AO simulation tool (COMPASS), we investigate the impact of several parameters on the performance of the AO system. We measure the performance in minimizing the star residuals in the coronagraphic image. The parameters that we study are : the second stage frequency, the photon flux on each WFS, the first stage gain and the DM number of actuators of the second stage. We show that the performance is improved by a factor 10 with respect to the current AO system (SAXO). The optimal second stage frequency is between 1 and 2 kHz under good observing conditions. In a red star case, the best SAXO+ performance is achieved with a low first stage gain of 0.05, which reduces the first stage rejection.
\end{abstract}


\section{INTRODUCTION}
\label{sec:intro}  

High-contrast imaging aims at detecting light emitted or reflected by the near surroundings of stars. This allows spectroscopic and polarimetric characterization of circumstellar disks and atmospheres of young giant exoplanets. Such observations are challenging because the star is much brighter than the exoplanets or the disk. That is why high-contrast instruments use a coronagraph to block the on axis starlight and let the off-axis light go to the detector. As coronagraphs are designed for a no-aberration wavefront, ground-based high-contrast instruments are assisted by an extreme adaptive optics (AO) system, such as GPI at Gemini South \cite{Macintosh_2018}, Clio2/Mag-AO at the Magellan telescope \cite{Sivanandam_2006, Close_2010}, SCExAO at Subaru \cite{Jovanovic_2015} and SPHERE at VLT \cite{Beuzit_2019}. The current raw detection limit of exoplanets is about $10^{-5}$ to $10^{-6}$ in terms of planet to star luminosity ratio, at a few hundreds mas.

SPHERE has been observing exoplanetary systems at the VLT since 2014. The SCAO of SPHERE, called SAXO \cite{Fusco_2014}, contains a 40*40 Shack-Hartmann (SH) wavefront sensor (WFS) working in visible light, a 41*41 high-order deformable mirror (HODM) and a fast tip-tilt mirror. Under high flux conditions, the loop frequency is 1.38 kHz. The AO system feeds the science instruments, equipped with visible and near-infrared coronagraphs. In good observing conditions, the intensity in the raw images, normalized by the maximum of the point spread function (PSF) with no coronagraph, goes down to $10^{-4}$ at 300 mas.

The SPHERE instrument is limited by uncorrected aberrations caused either by the AO residual turbulence or by optical aberrations of the telescope and the instrument. SAXO+ is a proposed upgrade of SAXO and is currently in the design phase \cite{Boccaletti_2020, Vidal_2023}. It aims at observing fainter and redder stars, and detecting more exoplanets, especially close to the star, between 100 and 300 mas. SAXO+ will tackle the current limitations of SAXO, namely the AO temporal error and the WFS sensitivity. To achieve these goals, SAXO+ includes a second stage AO, downstream of the actual SAXO stage. Its design is described in the next section. Finally, SAXO+ is part of the roadmap of the ESO instrument PCS/ELT \cite{Kasper_2021} as a technical demonstrator of a two-stage AO.

\section{SAXO+ design}

\begin{figure}[h]
    \centering
    \includegraphics[width=\textwidth]{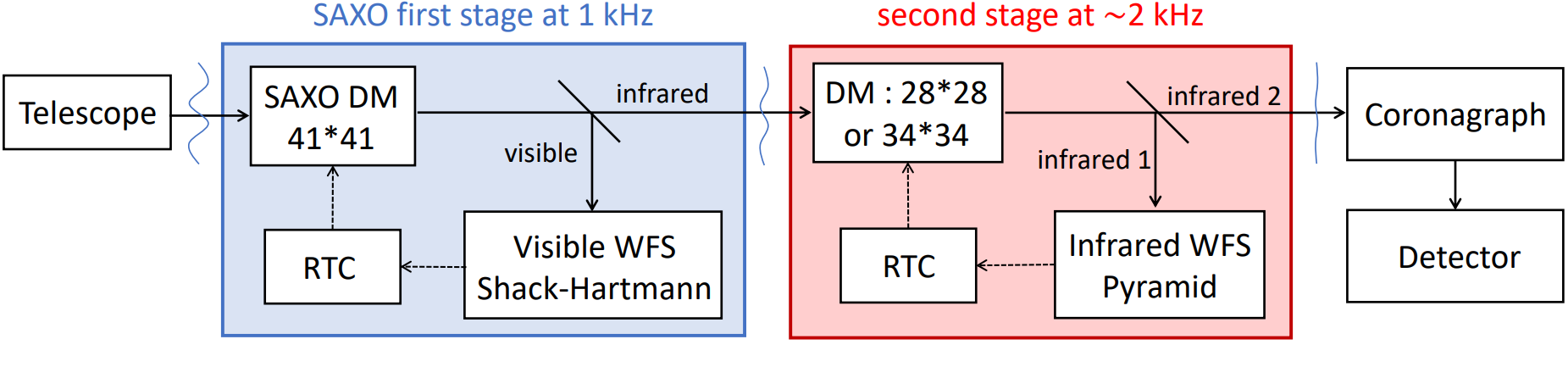}
    \caption{SAXO+ outline. In blue : current SAXO system. In red : second stage of SAXO+.}
    \label{design}
\end{figure}

The SAXO+ outline is described on the diagram figure \ref{design}. The first stage, in blue, is the current SAXO system. The second stage, in red, will run faster, between 1 and 3 kHz, to address the temporal error of the first stage. The wavefront sensing will be done in near-infrared with a pyramid WFS \cite{Ragazzoni_1996}, more sensitive than the SH WFS. This meets the science case of observing redder targets. In this design the two loops are independant (``stand alone'' case), with two separate real-time computer (RTC). An other possible solution consists in having only one RTC for the two WFSs and the two DMs (``integrated'' case). Then the pyramid WFS would also drive the first stage DM, and we could consider a more efficient control strategy. We will not explore such a solution in this paper.

One of the main system choices is the second stage frequency. Higher frequency implies theoritically better performance but higher technical constraints on the real-time subsystems (pyramid modulation mirror, RTC, ...). Then, there is a trade-off between the photon flux in the pyramid path and the science path. The cut-off wavelength will be between 1.2 and 1.4 µm, depending on the AO performance at low flux and the scientific requirements. Thirdly, two deformable mirrors are currently considered for the second stage, a 28*28 and 34*34 actuators. They correct lower orders than the HODM but are able to run faster, up to 3 kHz.

These system choices are critical and justify the needs of numerical simulations.

\section{Simulation implementation}

\subsection{Simulation tool}

To perform the simulations we used COMPASS, an end-to-end AO simulation tool \cite{Gratadour_2016, Compass_website} already used to design other AO systems (CANARY, MICADO, MAVIS, SCExAO, METIS, ...). As such, COMPASS had some limitations and we had to upgrade it to provide a comprehensive two-stage AO loop for SAXO+.

\subsubsection{Two-stage simulation}

\begin{figure}[h]
    \centering
    \includegraphics[width=0.9\textwidth]{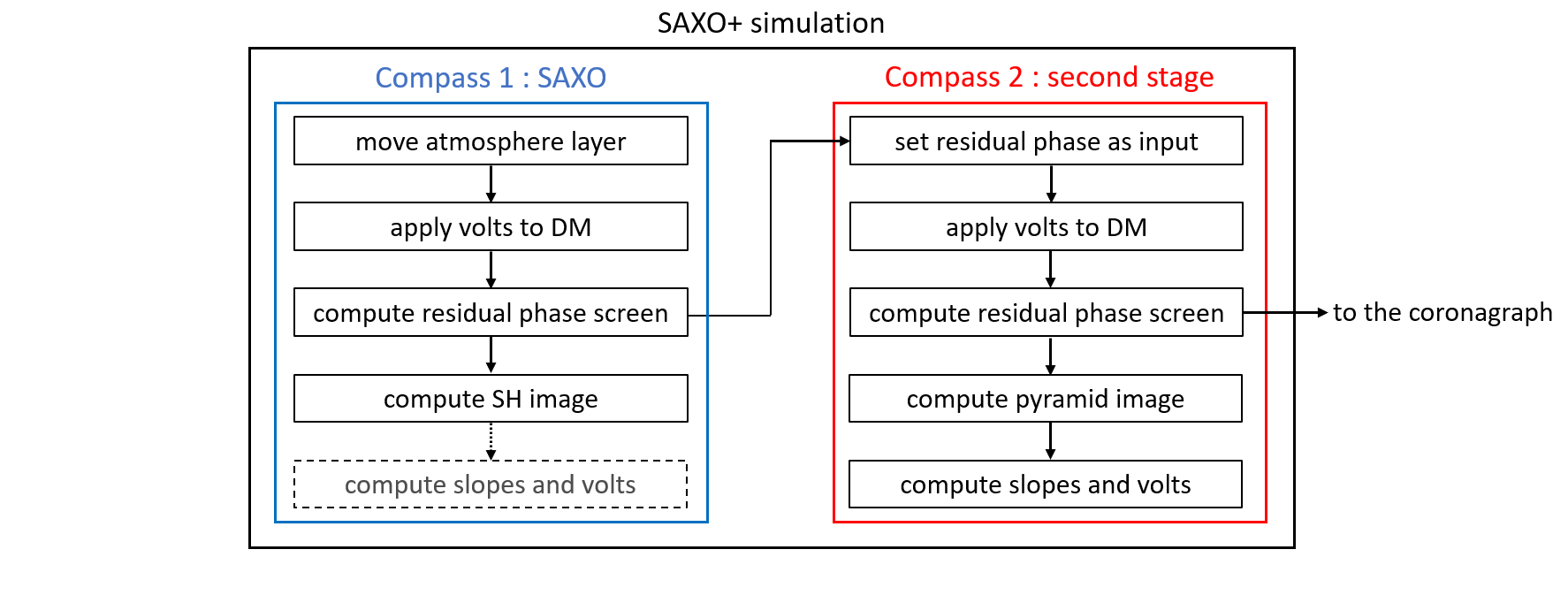}
    \caption{SAXO+ simulation structure. }
    \label{manager}
\end{figure}

Compass was limited to single frequency systems, and could not mix a SH WFS and a pyramid WFS in the same simulation. To simulate SAXO+, we created two Compass instances, one for SAXO and one for the second stage, hereafter referred to as Compass 1 and Compass 2 respectively. Figure \ref{manager} shows the steps carried out by each simulation in one iteration. In addition, Compass 1 is used to simulate the turbulent layers, while Compass 2 retrieves the residual phase screen of the first stage as an input to properly simulate a cascaded AO system.

In the real SAXO+ system, the first stage is running slower than the second stage. But in a SAXO+ simulation, the first stage has to be sampled at the same frequency (at least) than the second stage, because the second stage needs to see the turbulence moving at its frequency. To slow down the first stage correction, the step of computing the slopes and the command in volts (in dotted box figure \ref{manager}) is performed only one iteration over N, where N is the ratio between the second stage frequency and the first stage frequency. As N is an integer, we simulate systems where the second stage frequency is a multiple of the first stage frequency.

\subsubsection{Coronagraph module}

As a high-contrast instrument, the performance of SAXO+ must be assessed with residual intensity in coronagraphic images. We developped a coronagraph module for Compass \cite{Corono_tuto}. The coronagraph module takes the AO corrected phase screen as an input, simulates electric field propagation through the coronagraph and computes the coronagraphic image (meaning the distribution of the starlight on the science detector). Two types of coronagraphs are implemented : a realistic coronagraph and a perfect coronagraph.

The realistic coronagraph is based on the structure of the classical Lyot stellar coronagraph \cite{Paresce_Burrows_1986, Macintosh_1992, Nakajima_1994, Galicher_2023}.
It consists of four successive optical planes : a pupil apodizer, a focal plane mask, a Lyot stop and the image plane. An optical Fourier transform is performed to propagate the electric field from one plane to the next one. In the image plane, the square modulus of the electric field is computed to obtain the intensity.

The perfect coronagraph substracts the diffractive effect of the telescope \cite{Sauvage_phdthesis, Galicher_phdthesis}. Mathematically speaking, the following operations are performed:
\begin{enumerate}
    \setlength\itemsep{0em}
    \item Electric field is computed from the incoming phase screen
    \item The average of the electric field over the pupil is substracted to the electric field
    \item Optical Fourier transform and square modulus yield the image in focal plane
\end{enumerate}

The intensity in the coronagraphic image is normalized to the maximum of the AO corrected point spread function (PSF). For the realistic coronagraph, the PSF is computed without focal plane but with apodizer and Lyot stop. For the perfect coronagraph, the PSF is computed by removing the step number two. Such coronagraphic intensity is called normalized intensity hereafter.

\subsubsection{Graphical interface}

\begin{figure}[h]
    \centering
    \includegraphics[width=\textwidth]{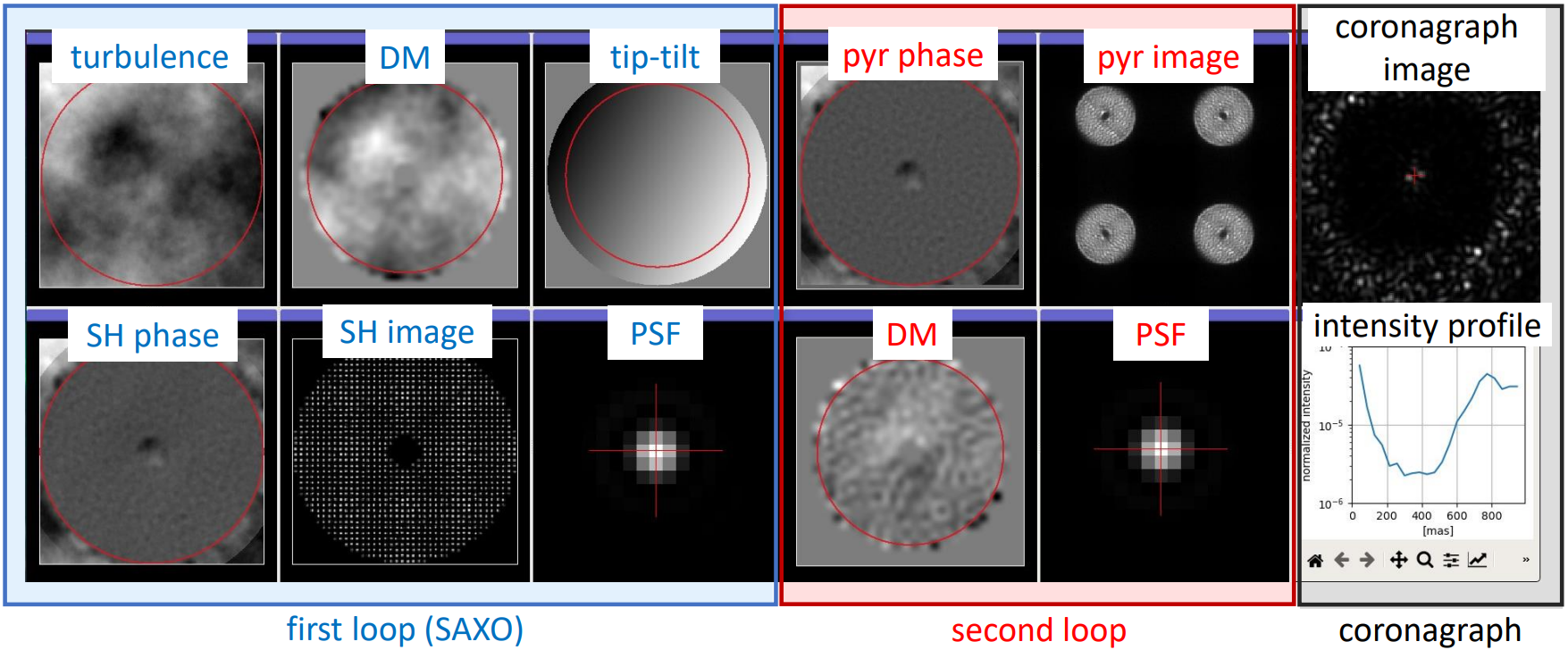}
    \caption{Compass graphical interface of a SAXO+ simulation. First stage in blue, second stage in red, and coronagraph in black. The first and second stage loops are closed.}
    \label{screen}
\end{figure}

Compass has a graphical display that allows the user to monitor critical subsystems. Figure \ref{screen} shows the graphical interface for a typical SAXO+ simulation. The first stage, SAXO is shown in blue. The top left corner shows the phase screen of the incoming turbulence. Then, there is the shape of the HODM and the tip-tilt mirror. Summing these three phases gives the residual turbulence at the bottom left corner. Using this residual phase screen, the SH image and the instantaneous PSF are computed.

The second stage, in red, takes the residual phase screen of the first stage as an input. Again, we can monitor the shape of the second stage DM, the residual pyramid phase,  the pyramid image and the PSF. Finally, the coronagraphic image and its intensity profile are shown in the two boxes on the right.

\subsubsection{Hardware implementation}

The Compass platform uses GPU acceleration. Our computer is equipped with 2 CPUs 20-core Intel$^{\mbox{\scriptsize{\textregistered}}}$ Xeon$^{\mbox{\scriptsize{\textregistered}}}$ E5-2698 v4 at 2.2 GHz and 512 GB RAM. The simulation were performed using 4 NVIDIA Tesla P100 16 GB GPUs. For SAXO+, the computation time is about 1 min per second of exposure.

\subsection{Simulation assumptions}

The baseline parameters are summarised in table 1. The first and second stage controllers are independent integrators. There is no compensation of the pyramid optical gains in the simulations from section 4.1 to 4.4. The CLOSE algorithm was used \cite{Deo_2021} in section 4.5. CLOSE estimates the optical gains in real time and compensates them with modal gains. We used a perfect coronagraph instead of one of the SPHERE coronagraphs so that we only have the light from the AO residuals in the coronagraphic images. This enable to study the AO performance with no limitation induced by the coronagraph. In section 4.2 to 4.5, we explore some specific parameters and discuss their impact on the system performance.

We emphasize that the results presented in this paper have been obtained under optimistic turbulence conditions (seeing and coherence time) and are not otherwise relevant. Such assumptions are incomplete for designing the final system and ongoing studies are currently performed with more pessimistic seeing and coherence times.

\begin{table}[h]
    \centering
    \caption{Baseline parameters. The parameters set in the rest of the article are shown in black. Variable parameters are shown in \textcolor{blue}{blue}, and their default value is given.}
    \begin{tabular}{|l|l|l|l|}
        \hline
        Observing conditions & First stage & Second stage & Coronagraph \\
        \hline
        8 m telescope & $\lambda_{\mathrm{WFS}}$ = 700 nm & $\lambda_{\mathrm{WFS}}$ = 1.2 \textmu m & Perfect coronagraph \\
        VLT pupil & 40*40 Shack-Hartmann & 50*50 pyramid & Imaging at $\lambda$ = 1.67 \textmu m \\
        1 atmospheric layer & \textcolor{blue}{High flux on WFS} & \textcolor{blue}{High flux on WFS} & 3 second exposure \\
        Seeing : 0.74" & Spatial filter size : 0.82" & 3 $\lambda / D$ modulation radius & No NCPA \\
        Coherence time : 5.5 ms & 41*41 DM, 800 KL modes & \textcolor{blue}{34*34 DM, 600 KL modes} & \\
        & Frequency : 1 kHz & \textcolor{blue}{Frequency : 2 kHz} & \\
        & \textcolor{blue}{Scalar gain : 0.4} & Scalar gain : 0.4 & \\
        \hline
    \end{tabular}
    \label{params}
\end{table}

\section{Results}

\subsection{SAXO versus SAXO+}

\begin{figure}[h!]
    \centering
    \includegraphics[width=\textwidth]{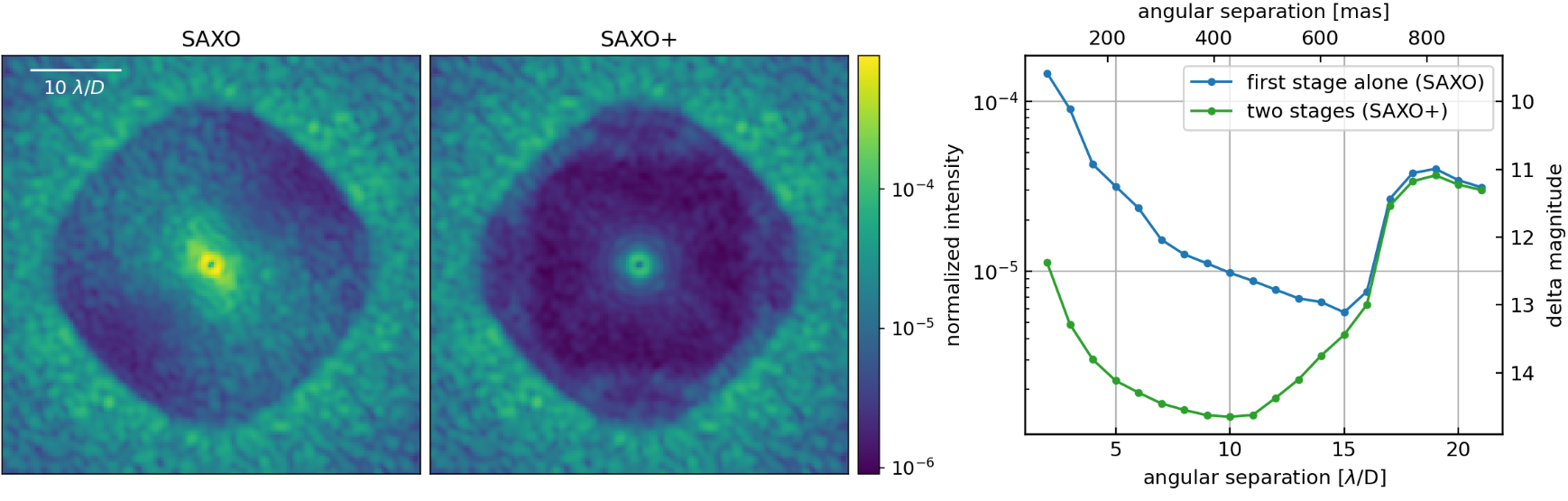}
    \caption{Left and center : Coronagraph images obtained after SAXO (left) and SAXO+ (center). Right : Corresponding normalized intensity curves.}
    \label{images}
\end{figure}

Thanks to COMPASS, we can compute coronagraph images with SAXO alone and after the two stages. An example is provided in figure \ref{images} on the left, with conditions described in table \ref{params}. In the case of SAXO alone, we can see the HODM correction area where the AO is effective. Within this region, the residual turbulence is distributed along a butterfly-shaped halo. This is due to the AO temporal error and is oriented in the direction of the wind. After the two stages (central image), the turbulence residuals are much lower within the second DM correction area.

To quantify the second stage improvement, we can compute the azimuthal mean intensity over 1 $\lambda / D$ width rings of various radii. These normalized intensity profiles are shown on figure \ref{images}, with the angular separation on the x-axis. At a 3 $\lambda / D$ separation, the normalized intensity after the first stage only is $9\cdot 10^{-5}$ and goes down to $5\cdot 10^{-6}$ with SAXO+. Between 2 and 10 $\lambda / D$, SAXO+ improves the performance by a factor 10 with respect to SAXO.

Such normalized intensity curves are the criteria we use in the following sections to quantify the system performance.

\subsection{Frequency of the second stage}

\begin{figure}[h]
    \centering
    \includegraphics[width=0.65\textwidth]{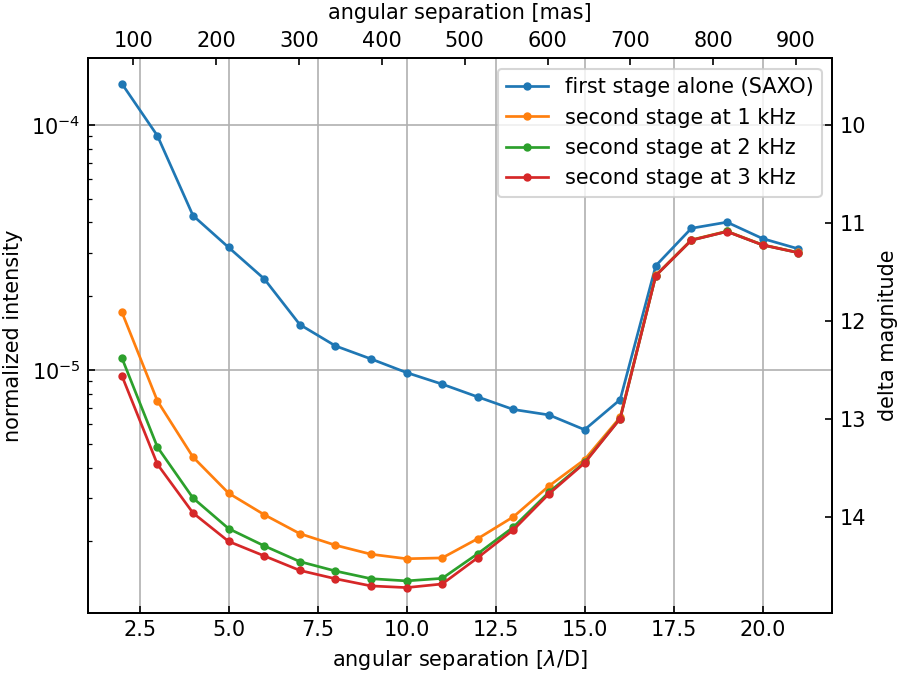}
    \caption{Normalized intensity profile after first stage alone and after the two stages, with second stage running at various frequencies.}
    \label{frequency}
\end{figure}

The choice of the second stage frequency is a trade-off between the SAXO+ performance and technical constraints on the real time system (pyramid modulation mirror, RTC, electronics…). We performed simulations to assess the performance gain as the second stage frequency increases. Figure \ref{frequency} represents normalized intensity profiles with the second stage running at 1 kHz, 2 kHz and 3 kHz. At a 3 $\lambda / D$ separation, the normalized intensity is $7.5\cdot 10^{-6}$ with the second stage running at 1 kHz, $4.9 \cdot 10^{-6}$ at 2 kHz and $4.1\cdot 10^{-6}$ at 3 kHz. The improvement of residual intensity is around 40 \% when going from 1 to 2 kHz. However, this improvement is no more significant when going from 2 to 3 kHz (less than 20 \%). Under the assumption of good observing conditions the optimal frequency is between 1 and 2 kHz. However, further studies simulating lower coherence times are needed to choose the best second stage frequency.

It is also relevant to notice that just by adding the second stage at the same speed than the first one, there is a huge improvement of performance, by a factor 10. Most of the improvement is related to the addition of a second-stage AO, no matter how much the frequency is increased. This is partly thanks to the use of a pyramid wavefront WFS, more sensitive than the SH. Again, this result can only be interpreted under optimistic seeing and coherence time.

\subsection{Photon flux on each WFS}

\begin{figure}[h]
    \centering
    \includegraphics[width=0.9\textwidth]{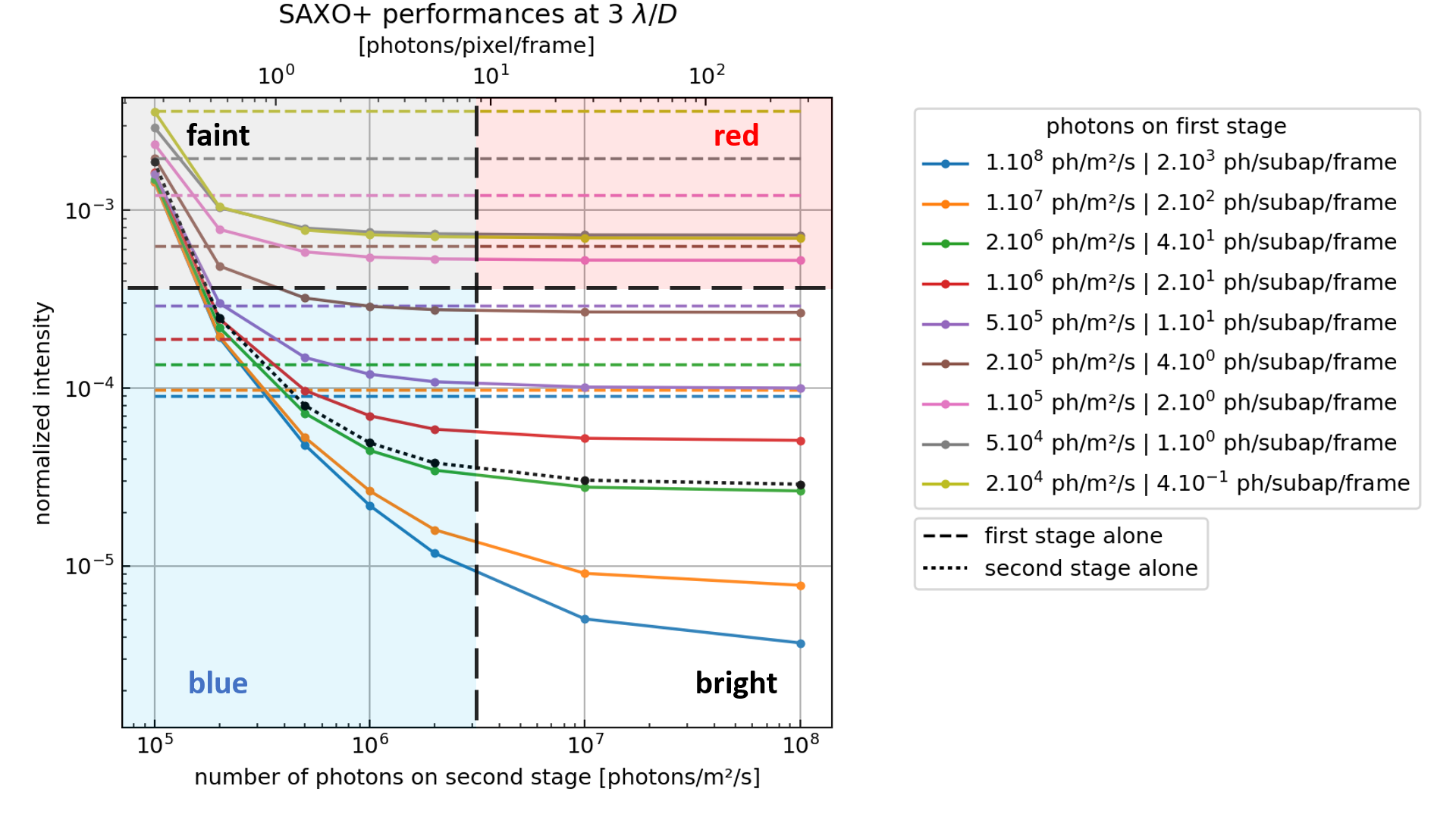}
    \caption{SAXO+ performance at 3 a $\lambda / D$ angular separation with respect to the photon flux on each WFS}
    \label{nb_photons}
\end{figure}

An other important system choice is the photon sharing between the pyramid path and the science path. Giving more photons to the pyramid WFS yields a better correction and a fainter limit magnitude but worse science signal to noise ratio (SNR). In figure \ref{nb_photons}, the normalized intensity at 3 $\lambda / D$ on the y-axis is investigated with respect to the number of photons on each WFS. The photon flux on the pyramid is on the x-axis, in photoelectrons/m$^2$/s for the bottom axis and in photoelectrons/pixel/frame for the top axis. Each curve color matches a photon flux on the SH WFS. The two stage curves are in solid lines, while the SAXO alone curves are in dashed lines. Finally, the black dotted curve corresponds to the second stage running alone, without the first stage.

The solid curves are always below the dashed curves, except for really low SNR on the pyramid WFS, below 0.3 photoelectrons/pixel/frame. This means that the second stage improves the first stage correction, even for faint targets. However, at 20 photons/subap/frame on the SH (red curve) and below, the second stage has better performance alone (black dotted curve) than with the first stage. It is mainly due to noise residuals of the first stage that are amplified by the second stage, decreasing the second stage performance and thus the overall performance. Further studies of this case is presented in the next section, optimizing the gain of the first stage.

This graph is useful to predict the SAXO+ performance as a function of the magnitude of the guide star, its color, and the choice of cut-off wavelength between the pyramid and science paths. For instance, each corner of the graph correspond to a type of star. In the bottom right corner, both pyramid and SH WFS have a lot of photons, this is a bright case. The top right corner is a red star case : the SH WFS has few photons while there is a lot of flux in the pyramid path. Conversely, the bottom left is a blue case, and the faint case is in the top left corner.

\subsection{Gain of the first stage}

\begin{figure}[h]
    \centering
    \includegraphics[width=0.6\textwidth]{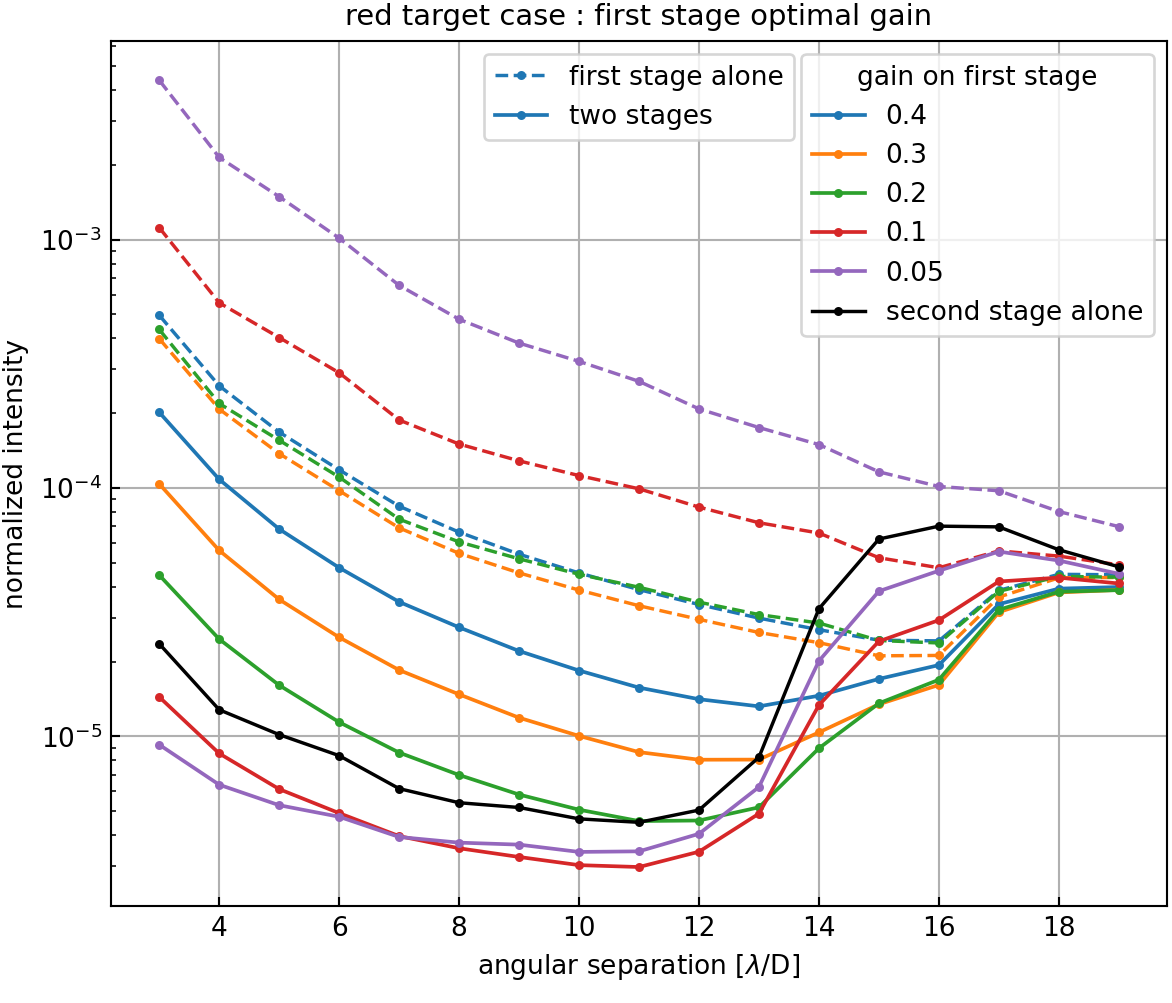}
    \caption{SAXO and SAXO+ performances for various gain on first stage. Red target case.}
    \label{optimal_gain}
\end{figure}

Red stars are one of the main science cases of SAXO+. It implies high flux on the pyramid WFS but low flux on the SH WFS. As we saw in the previous section, the first stage might have strong residuals, lowering the second stage correction. Figure \ref{optimal_gain} compares SAXO+ performance in a red target case (SH flux : 10 photons/subap/frame, pyramid flux : 1000 photons/pixel/frame) with respect to the first stage gain. The normalized intensity is represented on the y-axis versus the angular separation on the x-axis. First stage alone curves are in dashed lines while SAXO+ curves are in solid lines. Gains of the first stage are depicted by differents colors.

At a 3 $\lambda / D$ separation, the lowest normalized intensity for SAXO is $4 \cdot 10^{-4}$, obtained with a gain of 0.3 (dashed yellow curve). With the same gain, at 3 $\lambda / D$, the normalized intensity after SAXO+ is $10^{-4}$ (solid yellow curve). However, the best SAXO+ performance at 3 $\lambda / D$ is a normalized intensity of $9 \cdot 10^{-6}$, obtained with a gain of 0.05 on the first stage (solid purple curve). For the first stage alone, setting the gain to 0.05 yields a normalized of $4 \cdot 10^{-3}$ at 3 $\lambda / D$. We may infer that, in a red target case, slowing down the first stage by reducing its gain improves the two-stage performance in the region of interest. Similar results have been obtained in \cite{Cerpa-Urra_2022}.

\subsection{DM choice}

\begin{figure}[h]
    \centering
    \includegraphics[width=\textwidth]{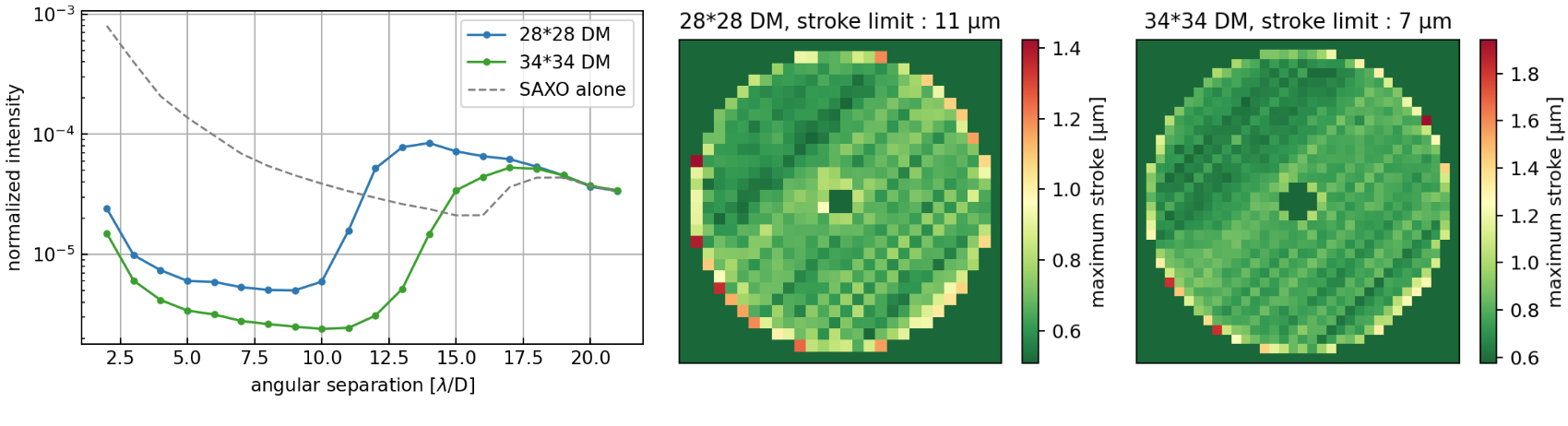}
    \caption{Comparison of 28*28 and 32*32 DM}
    \label{DM_choice}
\end{figure}

The DM of the second stage has to be fast enough to be commanded at 2 or 3 kHz. As SAXO+ aims at detecting exoplanets close to the star, this DM does not need many actuators. Considering also the cost of the DM, two mirrors have been selected : the 28*28 and 34*34 kilo-DMs from Boston Michromachines. The main difference between the two DMs is their stroke limitation. The 28*28 DM has a 11 µm optical stroke limit while the 34*34 has a shorter optical stroke limit of 7 µm. On figure \ref{DM_choice}, we compare the performances of the two DMs in the same red target case than the previous section. Following the previous section, the gain of the first stage has been set to 0.05 when the second stage is running, and to 0.3 when SAXO is running without the second stage.

On the left of figure \ref{DM_choice}, the normalized intensity with respect to the angular separation is represented for the two DMs. The SAXO alone curve in grey dashed line is shown as an landmark. At a 3 $\lambda / D$ separation, the SAXO performance is $4\cdot 10^{-4}$. After SAXO+ the normalized intensity is improved down to $10^{-5}$ with the 28*28 DM and $6\cdot 10^{-6}$ with the 34*34 DM. The SAXO+ performance for the two DMs is constant until 10 $\lambda / D$. Then we reach the edge of the 28*28 correction zone. At 13 $\lambda / D$, the normalized intensity with the 28*28 DM is $8\cdot 10^{-5}$, while the 34*34 performance is at $5\cdot 10^{-6}$. As expected, the 34*34 DM has a deeper and wider correction area than the 28*28. The normalized intensity at 15 $\lambda / D$ for SAXO alone is $2.1 \cdot 10^{-5}$, below the SAXO+ performance, which is $3.4 \cdot 10^{-5}$ with the 28*28 DM and $7.2 \cdot 10^{-5}$ with the 34*34 DM. This is because the separations between 14 and 16 $\lambda / D$ are inside the correction zone of the first stage but outside the correction zone of the 28*28 and 34*34 DMs, and the gain of the first stage is 0.3 for the SAXO alone curve,  while it is 0.05 when the second stage is running.

To the right of figure \ref{DM_choice} is the maximum displacement of each actuator during a 5 second exposure. The line pattern is due to the simulation of the turbulence, as we used only 1 atmospheric layer with a wind oriented at 45°. The maximum stroke reached by an actuator is 1.4 µm for the 28*28 DM and 1.9 µm for the 34*34 DM. Under the assumption of a 0.74" seeing, there is a 5 \textmu m margin for the 34*34 DM and more than 9 \textmu m for the 28*28. However, the stroke limitation needs to be adressed in the most pessimist seeing conditions. The remaining available stroke has to be larger than a given safety margin, considering that part of the stroke might be used to flatten the DM, correct for tip-tilt, common and non-common path aberrations, etc... The actuators highlighted in red exceeded a 1.2 µm stroke for the 28*28 DM and a 1.6 µm for the 34*34 DM. These actuators are located at the edge of the pupil. As a result, the maps show that the actuators on the pupil edge are the most likely to reach the stroke limit.

\section{Conclusion}

We studied the impact of main AO parameters on SAXO+ performances. In good observing conditions, the second stage improves by a factor 10 the residual star intensity in the coronagraph science image. The frequency 2 kHz seems to be reasonable for driving the second stage. For the red target case, first stage gain needs to be lowered (but not nulled) to get the best two stage performance. The 34*34 DM has a wider correction area than the 28*28, but is also closer to its stroke saturation. In the comming months, we will fully explore the parameter space to consolidate the SAXO+ design (seeing, coherence time of atmosphere and pyramid modulation amplitude).

\acknowledgments 
 
This work was supported by the Action Spécifique Haute Résolution Angulaire (ASHRA) of CNRS/INSU co-funded by CNES.


\printbibliography 
\end{document}